\documentclass[aps, twocolumn, superscriptaddress]{revtex4} 

\usepackage[switch]{lineno}
\usepackage{amsmath}
\usepackage{amssymb}
\usepackage{empheq}
\usepackage[parfill]{parskip}
\usepackage[active]{srcltx}
\usepackage{color}
\usepackage{array}
\usepackage{booktabs}
\usepackage{amsfonts}
\usepackage{dsfont}
\usepackage{graphicx}
\usepackage{appendix}
\usepackage{hyperref}
\usepackage{lineno}
\usepackage{enumerate}

\begin{document}
\setlength{\parindent}{0.5cm}
	
\title{Dynamics of swarmalators in the presence of a contrarian}
	
\author{Gourab Kumar Sar}
\email{mr.gksar@gmail.com}
\affiliation{Physics and Applied Mathematics Unit, Indian Statistical Institute, 203 B. T. Road, Kolkata 700108, India}
	
\author{Sheida Ansarinasab}
%\email{}
\affiliation{Department of Biomedical Engineering, Amirkabir University of Technology (Tehran Polytechnic), Iran}

\author{Fahimeh Nazarimehr}
\affiliation{Department of Biomedical Engineering, Amirkabir University of Technology (Tehran Polytechnic), Iran}

\author{Farnaz Ghassemi}
\affiliation{Department of Biomedical Engineering, Amirkabir University of Technology (Tehran Polytechnic), Iran} 
	
\author{Sajad Jafari}
\affiliation{Department of Biomedical Engineering, Amirkabir University of Technology (Tehran Polytechnic), Iran} \affiliation{Health Technology Research Institute, Amirkabir University of Technology (Tehran Polytechnic), Iran}
	
\author{Dibakar Ghosh}
\email{Corresponding to: diba.ghosh@gmail.com}
\affiliation{Physics and Applied Mathematics Unit, Indian Statistical Institute, 203 B. T. Road, Kolkata 700108, India}

\begin{abstract}
	\hspace{1 cm}  (Received XX MONTH XX; accepted XX MONTH XX; published XX MONTH XX) \\ \\
		Swarmalators are entities that combine the swarming behavior of particles with the oscillatory dynamics of coupled phase oscillators and represent a novel and rich area of study within the field of complex systems. Unlike traditional models that treat spatial movement and phase synchronization separately, swarmalators exhibit a unique coupling between their positions and internal phases, leading to emergent behaviors that include clustering, pattern formation, and the coexistence of synchronized and desynchronized states etc. This paper presents a comprehensive analysis of a two-dimensional swarmalator model in the presence of a predator-like agent that we call a contrarian. The positions and the phases of the swarmalators are influenced by the contrarian and we observe the emergence of intriguing collective states. We find that swarmalator phases are synchronized even with negative coupling strength when their interaction with the contrarian is comparatively strong. Through a combination of analytical methods and simulations, we demonstrate how varying these parameters can lead to transitions between different collective states.\\
	\noindent \\
	DOI: XXXXXXX
\end{abstract}
	
\maketitle
%%%%%%%%%%%%%%%%%%%%%%%%%%%%%%%%%%%%%
\section{Introduction}
\label{Section 1}
In recent years, the study of collective behavior in natural and artificial systems has gained significant attention, particularly in the realms of swarming and synchronization phenomena. These phenomena are observed in various contexts, from the flocking of birds and schooling of fish to the coordinated flashing of fireflies and the rhythmic beating of heart cells~\cite{sumpter2010collective,rosenblum2003synchronization}. Traditionally, swarming and synchronization have been studied separately, with models focusing either on the spatial dynamics of swarms or the temporal dynamics of coupled oscillators~\cite{okubo1986dynamical,acebron2005kuramoto,sar2023flocking}. However, the advent of the swarmalator model has provided a novel framework that integrates these two domains, capturing the interplay between an agent's spatial movement and its internal phase dynamics. Swarmalators, a portmanteau of ``swarm" and ``oscillator", are entities that simultaneously exhibit swarming behavior and oscillatory dynamics, leading to a rich variety of emergent phenomena that cannot be explained by swarming or synchronization alone~\cite{o2017oscillators,sar2022dynamics,o2019review}. The study of swarmalators is not only theoretically compelling but also has practical implications for understanding natural systems, such as the collective dynamics of biological organisms~\cite{aihara2014spatio}, and for designing artificial systems in robotics and networked technologies~\cite{barcis2020sandsbots}.

In 2017, O'Keeffe et al.~\cite{o2017oscillators} first proposed a model of swarmalators moving in a two-dimensional (2D) plane. They considered $N$ swarmalators moving in 2D with position $\mathbf{x}_i = (x_i,y_i)$ and an internal phase $\theta_i$. The positions and phases have a mutual influence on each other and their governing equations are
\begin{align}
	\Dot{\mathbf{x}}_i=& \dfrac{1}{N}\sum_{j \ne i}\Bigg[(\mathbf{x}_j-\mathbf{x}_i) \big(1 + J \cos(\theta_j - \theta_i)\big)- \dfrac{\mathbf{x}_j-\mathbf{x}_i}{|\mathbf{x}_j-\mathbf{x}_i|^2}\Bigg],
	\label{eq1}\\
	\Dot{\theta}_i=& \dfrac{K}{N}\sum_{j \ne i} \dfrac{\sin(\theta_j-\theta_i)}{|\mathbf{x}_j-\mathbf{x}_i|},
	\label{eq2}
\end{align}
for $i=1,2,...,N$. There are spatial attraction and repulsion between the $i$-th and the $j$-th swarmalator that are given by $\mathbf{x}_j-\mathbf{x}_i$ and $(\mathbf{x}_j-\mathbf{x}_i)/|\mathbf{x}_j-\mathbf{x}_i|^2$, respectively. Both of these are necessary for the boundedness and the collision avoidance of the solutions. The spatial attraction is influenced by the phase difference through the term $1 + J \cos(\theta_j - \theta_i)$. When $J>0$, swarmalators in nearby phase attract themselves spatially. The phase equation~\eqref{eq2} is inspired from the Kuramoto model~\cite{10.1007/BFb0013365} and is further influenced by the spatial distance between the swarmalators where $K$ is the phase coupling strength. This model has been studied extensively over the last few years both numerically and theoretically~\cite{o2018ring,ha2019emergent,ceron2023diverse,ansarinasab2024spatial,ansarinasab2024thespatial,ghosh2024amplitude}. Many variations of the model have been explored by researchers by introducing external forcing~\cite{lizarraga2020synchronization}, different coupling mechanisms~\cite{hong2021coupling,lee2021collective,sar2022swarmalators}, various network structures~\cite{ghosh2023antiphase,lizarraga2024order,kongni2023phase} etc. Swarmalators have also been positioned on a one-dimensional (1D) ring and their collective behaviors have been studied~\cite{o2022collective,yoon2022sync,o2022swarmalators,sar2023pinning,sar2023swarmalators}.

This paper sets out to study the dynamics of swarmalators in the 2D plane under the influence of a predator-like agent. An external agent impacts the spatio-temporal properties of the swarmalator system as it interacts with each swarmalator both spatially and in phase. Earlier, the dynamics of swarmalators under an external periodic forcing were reported in Ref.~\cite{lizarraga2020synchronization}. To the best of our knowledge, the spatial component has been kept unaltered in most of these studies. In the presence of a predator-like agent, both the spatial and the phase dynamics of the swarmalators are influenced in our study.

In general, the introduction of a predator or an external agent can have significant and complex effects on the overall dynamics of a multi-agent system. The presence of a predator often induces a range of adaptive behaviors among the agents, such as increased cooperation, strategic positioning, or the development of defensive mechanisms~\cite{milinski1997cooperation,evans1990insect}. These behavioral changes can lead to emergent properties within the system, such as the formation of clusters or the development of new communication strategies among agents to mitigate the predator's impact~\cite{abrams2000evolution}. The predator's influence may also drive the system towards a new equilibrium, where the agents' interactions and strategies are continually optimized in response to the threat. This dynamic interplay between the agents and the predator highlights the system's robustness, adaptability, and potential for self-organization in the face of external pressures. In our case too, we find that swarmalators synchronize their phases even with negative phase coupling ($K<0$ in Eq. \eqref{eq2}) when their connection with the external agent is sufficiently strong. We also find new states that emerge due to the interaction with the external agent, like predators.

The rest of the paper is organized as follows. In Sec.~\ref{model}, we define our model of swarmalators with contrarians and discuss several aspects of it. In Sec.~\ref{single}, we study our model with a single contrarian. We investigate the spatial and phase structures with varying coupling strengths. The effects of these coupling strengths are discussed in detail. We provide analytical support to our numerical findings by solving the radii of the annular sync and async states. In Sec.~\ref{mult}, we briefly mention some results of our model with more than one contrarian. Finally, we discuss the findings of our study along with mentioning possible directions of future works in Sec.~\ref{discussion}.

\section{Proposed model}\label{model}
Our goal is to study the swarmalator system~\eqref{eq1}-\eqref{eq2} in the presence of some predator-like agents that have a paramount impact on the long-term dynamics of the system. Like the swarmalators, these agents also have both spatial and phase dynamics. We call these agents \textbf{contrarians}, as their dynamics are different from those of regular swarmalators. Let $p$ be the number of contrarians, and $\mathbf{z}_i$, $\psi_i$ denote the positions and phases, respectively, for $i=1,2,\ldots,p$. The number of contrarians is considered to be very small compared to the number of swarmalators, i.e., $p \ll N$ (the results of the model remain unchanged till this condition is satisfied). Now, consider the model below
\begin{align}
    \Dot{\mathbf{x}}_i=& \dfrac{1}{N}\sum_{j \ne i}\Bigg[(\mathbf{x}_j-\mathbf{x}_i) \big(1 + J \cos(\theta_j - \theta_i)\big)- \dfrac{\mathbf{x}_j-\mathbf{x}_i}{|\mathbf{x}_j-\mathbf{x}_i|^2}\Bigg] \nonumber \\ &+ \dfrac{1}{p}\sum_{j =1}^{p}\Bigg[ (\mathbf{z}_j-\mathbf{x}_i) -  \dfrac{\mathbf{z}_j-\mathbf{x}_i}{|\mathbf{z}_j-\mathbf{x}_i|^2}\Bigg],
    \label{eq3}\\
    \Dot{\theta}_i=& \dfrac{K_1}{N}\sum_{j \ne i} \dfrac{\sin(\theta_j-\theta_i)}{|\mathbf{x}_j-\mathbf{x}_i|}  + \dfrac{K_2}{p}\sum_{j =1}^{p} \dfrac{\sin(\psi_j-\theta_i)}{|\mathbf{z}_j-\mathbf{x}_i|}.
    \label{eq4}
\end{align}
There are spatial attraction and repulsion between the swarmalators and the contrarians which are expressed by the terms inside the second summation in Eq.~\eqref{eq3}. Swarmalator phases interact with the contrarians' phases with coupling strength $K_2$ and the coupling is affected by the spatial distance between them. In our study, the contrarians' positions and phases are considered to be fixed and  we take them as
\begin{equation}
    \mathbf{z}_i = \left(r_i\cos \dfrac{2 \pi (i-1)}{p},r_i\sin \dfrac{2 \pi (i-1)}{p}\right),\hspace{1 pt}
    \psi_i = \dfrac{2 \pi (i-1)}{p},
    \label{contrarian}
\end{equation}
for $i=1,2,...,p$, where $r_i$ is the spatial distance of the $i$-th contrarian from the origin.

{\bf Order parameters:}
We define the following order parameter
\begin{equation}
    Re^{i\Phi} = \frac{1}{N} \sum_{j=1}^{N} e^{i\theta_j},
\end{equation}
which is the usual Kuramoto order parameter that measures the coherence among swarmalators' phases. We also define 
\begin{equation}
	S_{\pm}e^{i\Psi_{\pm}} = \frac{1}{N} \sum_{j=1}^{N} e^{i(\phi_j \pm \theta_j)},
\end{equation}
where $\phi=\tan^{-1}(y/x)$ is the spatial angle of the swarmalators. $S_{\pm}$ measure the correlation between swarmalators' spatial angles and phases. By their definition, $R$ and $S_{\pm}$ always lie between $0$ and $1$. $R=1$ means that swarmalators' phases are completely synchronized. Similarly, $S_{\pm}=1$ means the spatial angles of the swarmalators ($\phi$) and the phases ($\theta$) are fully correlated.

\begin{figure*}[hpt]
	\centering
	\includegraphics[width=1.6\columnwidth]{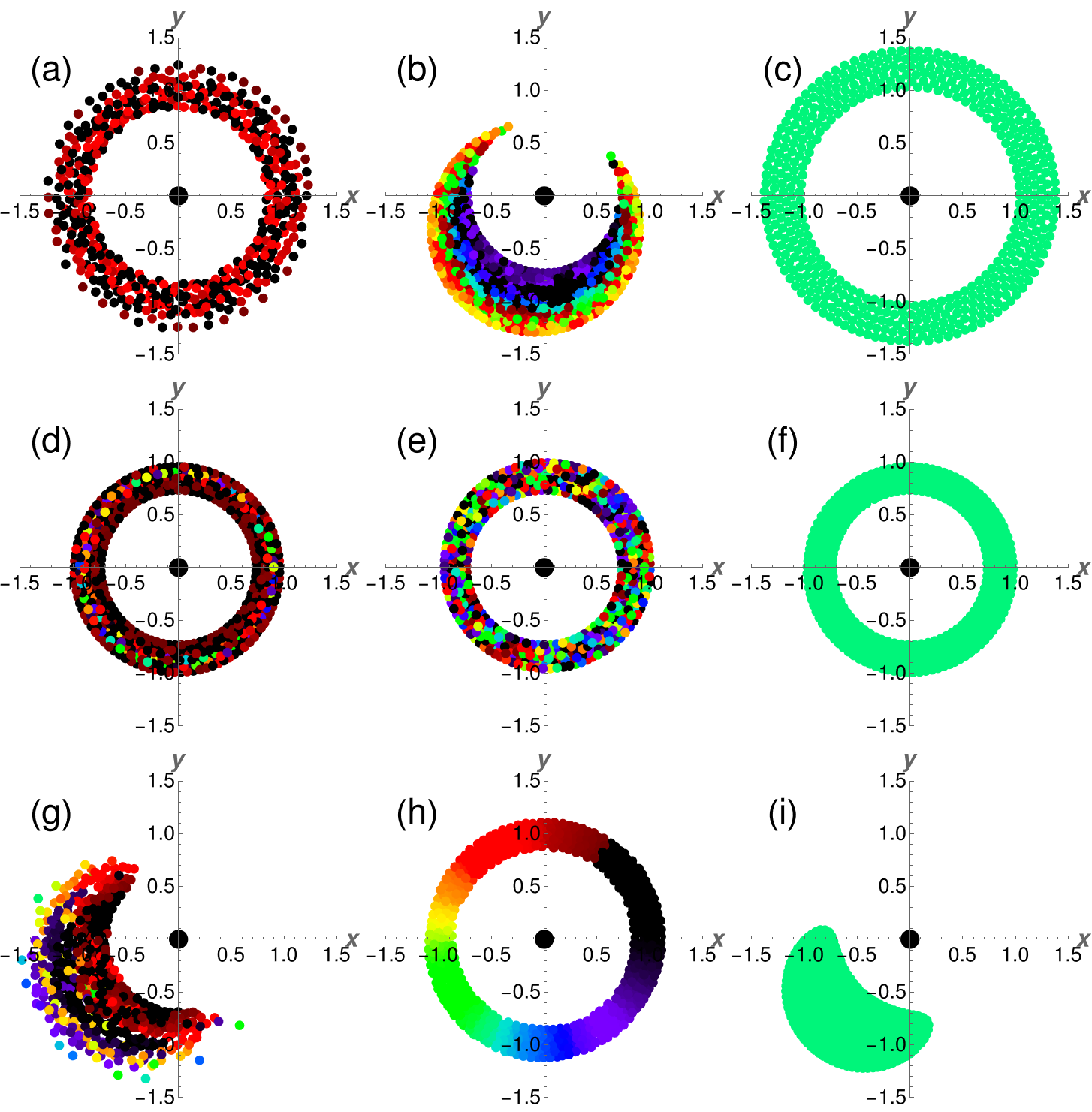}
	\caption{The collective behaviors of swarmalators on the $x$-$y$ plane. The first row (a-c), second row (d-f), and third row (g-i) are for $J=-1$, 0, and 1, respectively. The coupling parameters are $K_1=-K_2=-1$, $K_1=-K_2=0$, and $K_1=-K_2=1$ in the first, second and third columns, respectively. In these diagrams, the color assigned to each swarmalator represents its phase in the interval $[0,2\pi]$. The swarmalators' initial positions and phases are randomly chosen within the intervals $[-1,1] \times [-1,1]$ and $[0,2\pi]$, respectively. The contrarian is positioned at a fixed location of $[0,0]$ with a phase of $0$ and is indicated by the big black dot at the origin. The simulations are performed with a runtime of $T=1000$, a time step of $dt=0.1$, and involve a total of $N=500$ swarmalators.} 
	\label{states}
\end{figure*}

\section{Single contrarian ($p=1$)}\label{single}
First, we focus on the swarmalator dynamics in the presence of a single contrarian. The model \eqref{eq3}-\eqref{eq4} for a single contrarian $p=1$ boils down to
\begin{align}
    \Dot{\mathbf{x}}_i=& \dfrac{1}{N}\sum_{j \ne i}\Bigg[(\mathbf{x}_j-\mathbf{x}_i) \big(1 + J \cos(\theta_j - \theta_i)\big)- \dfrac{\mathbf{x}_j-\mathbf{x}_i}{|\mathbf{x}_j-\mathbf{x}_i|^2}\Bigg]\nonumber \\ &+ (\mathbf{z}_1-\mathbf{x}_i) - \dfrac{\mathbf{z}_1-\mathbf{x}_i}{|\mathbf{z}_1-\mathbf{x}_i|^2},
    \label{xdot}\\
    \Dot{\theta}_i=& \dfrac{K_1}{N}\sum_{j \ne i} \dfrac{\sin(\theta_j-\theta_i)}{|\mathbf{x}_j-\mathbf{x}_i|} + K_2 \dfrac{\sin(\psi_1-\theta_i)}{|\mathbf{z}_1-\mathbf{x}_i|}.
    \label{thetadot}
\end{align}
From Eq.~\eqref{contrarian}, we extract the contrarian's position and phase as $\mathbf{z}_1 = (r_1,0)$ and $\psi_1 = 0$. However, the overall dynamics of the system remains qualitatively the same regardless of the position and the phase of the contrarian. In the 2D plane, the swarmalator dynamics is always centered near the position of the contrarian. Without loss of generality, we choose $\mathbf{z}_1 = (0,0)$ and $\psi_1 = 0$ which further simplify our model to
\begin{align}
    \Dot{\mathbf{x}}_i=& \dfrac{1}{N}\sum_{j \ne i}\Bigg[(\mathbf{x}_j-\mathbf{x}_i) \big(1 + J \cos(\theta_j - \theta_i)\big)- \dfrac{\mathbf{x}_j-\mathbf{x}_i}{|\mathbf{x}_j-\mathbf{x}_i|^2}\Bigg] \nonumber \\ &- \mathbf{x}_i +  \dfrac{\mathbf{x}_i}{|\mathbf{x}_i|^2},
    \label{space}\\
    \Dot{\theta}_i=& \dfrac{K_1}{N}\sum_{j \ne i} \dfrac{\sin(\theta_j-\theta_i)}{|\mathbf{x}_j-\mathbf{x}_i|} - K_2 \dfrac{\sin\theta_i}{|\mathbf{x}_i|}.
    \label{phase}
\end{align}
 The phase equation (Eq.~\eqref{phase}) is a generalization of coupled oscillator systems under an external field where the phases are influenced by the spatial distance~\cite{shinomoto1986phase,sakaguchi1988cooperative}.
There are two components in the phase equation \eqref{phase}. The first term controls the entrainment among the swarmalators' phases. The swarmalators' phases are entrained or synchronized when they are coupled with a positive (attractive) coupling strength, i.e., $K_1 >0$. The second term stands for the relation of the swarmalators' phases with that of the contrarian's. A positive value of $K_2$ reduces the difference between swarmalators' and contrarian's phases. The emerging behavior depends on the values of $J$, $K_1$ and $K_2$.

\begin{figure*}[hpt]
	\centering
	\includegraphics[width=2\columnwidth]{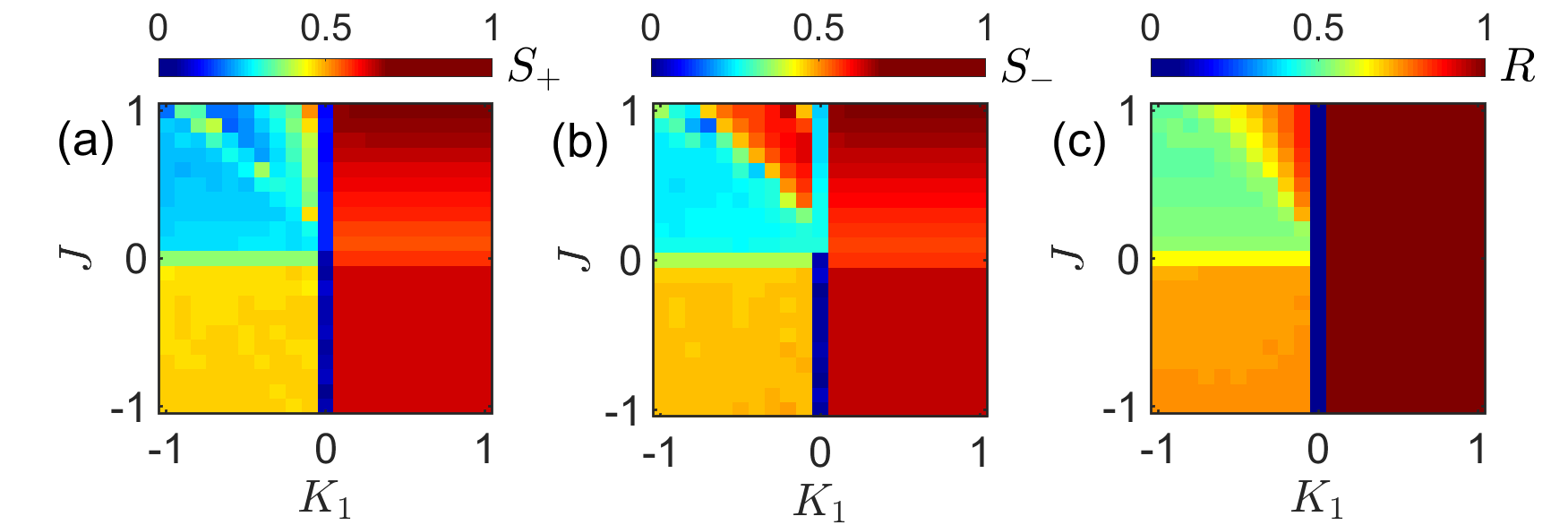}
	\caption{Diagrams of order parameters (a) $S_{+}$, (b) $S_{-}$ and (c) $R$ in the $K_{1}-J$ parameters plane. In these figures, the horizontal and vertical axes show the changes in phase coupling strength $K_{1}=-K_{2}$ and parameter $J$, respectively. In these matrices, the color assigned to each entry according to the colorbar indicates the intensity of the corresponding order parameter. The swarmalators' initial positions and phases are randomly chosen within the intervals of $[-1,1]$ and $[0,2\pi]$, respectively. The contrarian is positioned at a fixed location of $(0,0)$ with a phase zero. The simulations are performed with a runtime of $T=1000$, a time step of $dt=0.1$, and involve a total of $N=500$ swarmalators. In each simulation, $90\%$ of the total samples are discarded as transition, and the averages of the order parameters are calculated based on the remaining samples.} 
	\label{par-space}
\end{figure*}

{\bf Numerics:} Initially, the positions of the swarmalators are drawn from the box $[-1,1] \times [-1,1]$. The spatial dynamics guarantees that the center of positions of the swarmalators is conserved over time, a result known from Refs.~\cite{ha2019emergent,sar2022swarmalators}. Choice of a different range of the initial conditions just shifts the center of positions for the long-term dynamics of the emerging states, keeping their asymptotic behaviors unaffected. Swarmalators' initial phases are chosen from the interval $[0,2\pi]$. For simulations of the model, we numerically integrate Eqs.~\eqref{space}-\eqref{phase} by using the RK4 method with a step-size $0.1$ in the Mathematica and MATLAB softwares. The number of swarmalators $N$ has been chosen 500 for this study. Other simulation parameters and methods have been mentioned whenever required.

\subsection{Opposing phase interaction, $K_1 = - K_2$}
In this section, the findings derived from the simulation of Eqs.~\eqref{space}-\eqref{phase} pertaining to the scenario of opposing phase coupling strengths, denoted as $K_{1}=-K_{2}$, are presented. Figure~\ref{states} illustrates the swarmalators' collective behaviors on the $x$-$y$ plane across various sets of parameters $J$ and $K_{1}=-K_{2}$. In all the subplots, the big black dot at the origin indicates the position of the contrarian while the color (black) corresponds to its phase ($\psi_1=0$). These diagrams demonstrate that under a fixed value of parameter $J$, adjustments to the phase coupling strengths $K_{1}=-K_{2}$ for the cases of $K_{1}=0$, $0<K_{1}\le 1$, and $-1\le K_{1}<0$ lead to distinct alterations in the collective behaviors of swarmalators in both phase and spatial domains. For the case of $J=0$, it is observed that the contrarian is enclosed by a ring formed by the swarmalators. Conversely, for instances where $0<J \le 1$ and $0<K_{1} \le 1$, a distinctive behavior emerges where the contrarian exists beyond the spatial confinement of the swarmalators. Furthermore, Figs.~\ref{states}(c), (f), and (i) indicate that in the scenario of $0<K_{1}$, irrespective of the parameter $J$ value, the swarmalators exhibit synchronous phases with a phase difference of $\pi\hspace{0.08cm}$ radian relative to the contrarian. Noteworthy is the observation that the swarmalators' collective behaviors under conditions of $0<J \le 1$ and $K_{1}=-K_{2}=0$ mirror those governed by Eqs.~\eqref{eq1} and~\eqref{eq2} for $K=0$. Known as the ``static phase wave" state reported in ~\cite{o2017oscillators,ansarinasab2024spatial}, this configuration entails the maintenance of swarmalators' phases at their initial values over time. The positive parameter $J$ facilitates the aggregation of swarmalators sharing phases close to each other. We will discuss the effect of $J$ on the emerging states in details in Sec.~\ref{effect-j}.

The diagrams of order parameters $S_{\pm}$ and $R$ depicted in Fig.~\ref{par-space} provide insights into the distinction of swarmalators' collective behaviors resulting from variations in parameters $J$, $K_{1}$, and $K_{2}$. For each specific value of the $J$, when the phase coupling strength $0<K_{1} \le 1$, the order parameters $S_{\pm}$ exhibit high values. Concurrently, the order parameter $R$ shows a value of one, indicating phase synchronization among the swarmalators. When the phase coupling strengths are set to $K_{1}=-K_{2}=0$, the order parameters $S_{\pm}$ become smaller, and the order parameter $R$ descends to its minimum, underscoring the essential role of phase coupling strengths in driving the phase synchronization between the swarmalators. Additionally, these diagrams highlight a region within the parameter space—specifically $0.8<J\le 1$ and $-0.3<K_{1}<-0.1$—where the order parameters $S_{+}$ and $S_{-}$ show small and large values, respectively. This pattern signifies a correlation between the swarmalators' spatial arrangements and phases.

\iffalse
According to Fig.~\ref{cluster}, in these parameters, swarmalators sharing identical phases tend to aggregate near, exhibiting clustering-like behavior.
\begin{figure}[hpt]
	\centering
	\includegraphics[width=0.6\columnwidth]{fig3.png}
	\caption{The clustering-like collective behavior of the swarmalators in the $x$-$y$ plane for the parameter values of $J=0.9$, $K_{1}=-0.1$ and $K_{2}=0.1$. In this diagram, the color assigned to each swarmalator represents its phase. The simulations are performed with a runtime of $T=1000$, a time step of $dt=0.1$, and involve a total of $N=500$ swarmalators. Swarmalators show the same clustering-like collective behavior in other related parameter sets.} 
	\label{cluster}
\end{figure}
\fi

\subsection{Effect of $K_2$}
Every swarmalator's phase is coupled to that of the constant phase of the contrarian with coupling strength $K_2$. Swarmalators' phases are also coupled among themselves with coupling strength $K_1$. Simply put, a positive value of $K_1$ tries to minimize the phase difference among the swarmalators, whereas, a positive value of $K_2$ minimizes the phase difference between the swarmalator and the contrarian. When $K_1>0$, the swarmalators' phases are completely synchronized regardless of the value of $K_2$. Depending on whether $K_2$ is positive or negative, there exist two cases when $K_1>0$. If $K_2>0$ then the phases of all the swarmalators will be same as the contrarian's phase which is $\psi_1=0$ in our case. On the other hand, if $K_2<0$, then all the swarmalators' phases will maintain $\pi$ difference from the contrarian's phase, and in our case, they will be equal to $\pi$.

\begin{figure}[t]
\includegraphics[width=\columnwidth]{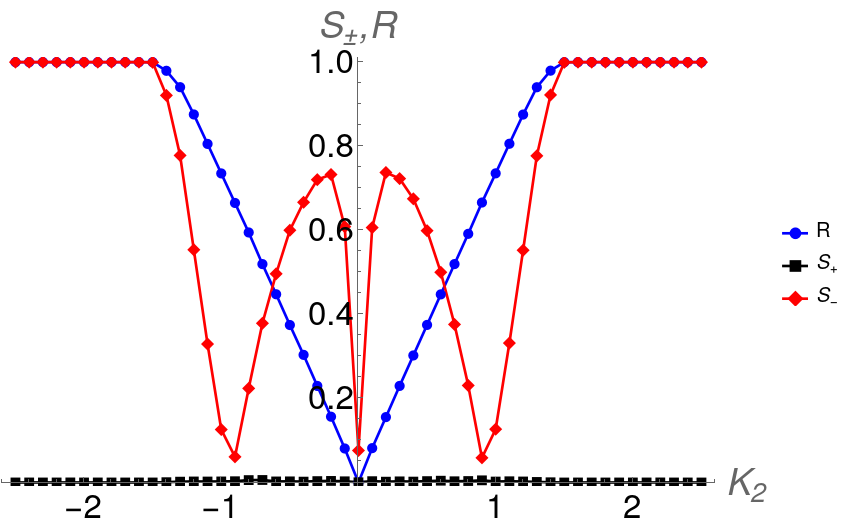}
\caption{Order parameters $R$, $S_+$, and $S_-$ as functions of $K_2$. For simulations, we take $N=500$, $T=1000$, $dt=0.1$. Here, $J=-0.5, K_1=-1$. Blue dots, black squares, and red diamonds represent the numerically calculated values of $R$, $S_+$, and $S_-$, respectively. Each data is an average of $10$ different initial conditions where the last $10\%$ data has been used after discarding the transients.}
\label{vary-k2}
\end{figure}

The more intriguing scenario, however, takes place when $K_1<0$. This means swarmalators' phases cannot achieve synchrony by themselves overcoming their phase differences. Synchrony can be achieved if the value of $K_2$ is positive and large enough so that each of the swarmalator's phase is synchronized to the contrarian's phase ($\psi_1=0$), and eventually the swarmalators are phase synchronized among themselves and also with the contrarian. At this point, $R$ becomes $1$. We denote this critical value as $K_2^* (>0)$. The other scenario, where $K_2$ is negative and large in magnitude is also viable for achieving synchrony among swarmalators' phases. Here, the magnitude of $K_2$ is negatively large enough that all the swarmalators' phases are in $\pi$ difference with the contrarian's phase. This results in complete phase synchronization among swarmalators which is at a difference of $\pi$ from $\psi_1=0$. We have found numerically that it happens when $K_2\le-K_2^*$. The key feature to notice here is that complete synchronization among swarmalators' phases can be observed even with negative phase coupling strength $K_1$ and the contrarian plays the pivotal role here. In Fig.~\ref{vary-k2}, we have fixed $J=-0.5, K_1=-1$ and varied $K_2$ and demonstrate the variation of the order parameters $R$, $S_+$, and $S_-$. The blue dots indicate the values of sync order parameter $R$ and we find from the figure that $R$ approaches $1$ when $|K_2|>K_2^*\approx1.5$. We also see that $S_+$ (black squares) always lies close to zero and $S_-$ (red diamonds) shows some non-monotonic behavior that arises from the spatial structure of the swarmalators which we discuss later. The other noticeable thing is the symmetry of the $R, S_+$ and $S_-$ graphs around $K_2=0$. For a fixed $K_1$, if we vary $K_2$, then the swarmalator dynamics are qualitatively the same for the values $\pm K_2$. From Eq.~\eqref{thetadot}, we acknowledge the fact that $-K_2$ value here corresponds to $K_2$ value if we choose the contrarians phase $\psi_1=0$. So, the symmetry in Fig.~\ref{vary-k2} also validates our earlier claim that the dynamics of the swarmalators remain unaltered regardless the phase of the contrarian.

\begin{figure}[t]
\includegraphics[width=\columnwidth]{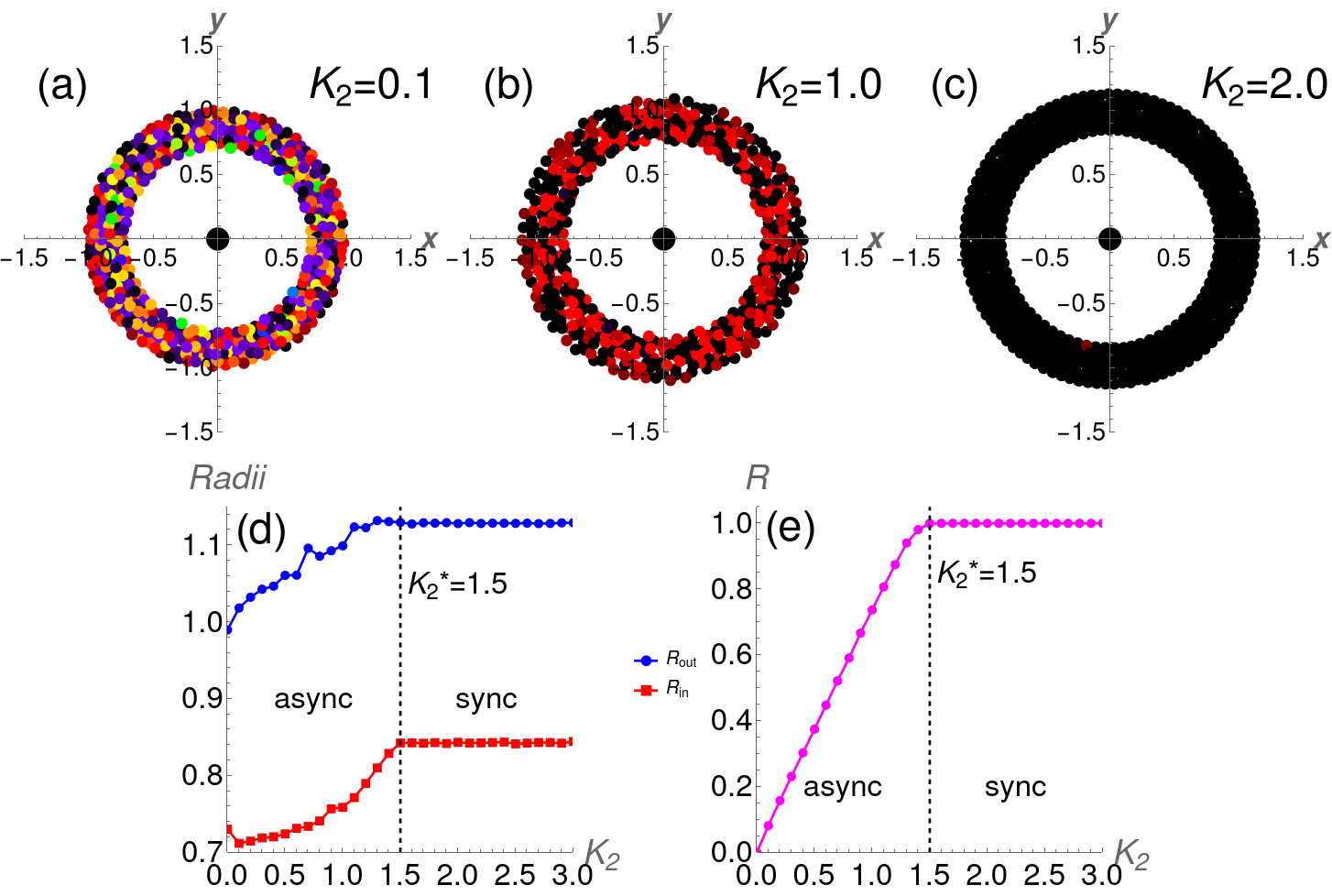}
\caption{Variation of the annular state depending on $K_2$. Here we choose $J=-0.5, K_1=-1.0$. The annular states are shown for (a) $K_2=0.1$, (b) $K_2=1.0$, and (c) $K_2=2.0$. The variations of the outer ($R_{out}$) and inner ($R_{in}$) radii are illustrated in (d) through blue dots and red square markers, respectively. The change of $R$ is highlighted in (e). For simulations we have used $(dt,T,N)=(0.1,1000,500)$.}
\label{annular-radii}
\end{figure}

Continuing the study with $J=-0.5,K_1=-1$, now we delve into the spatial structure of the emerging states and explore the affect of $K_2$ on them. With these choices of $J$ and $K_1$, swarmalators always seem to spatially arrange themselves inside an annular structure in 2D. Their phase pattern changes as the coupling strength $K_2$ varies. When $K_2$ is near $0$, the phases are desynchronized and the value of $R$ is near $0$. See Fig.~\ref{annular-radii} (a) for the spatial structure of the emerging state at $K_2=0.1$. As the value of $K_2$ increases, the phases become more coherent which can be seen in Fig.~\ref{annular-radii} (b) where $K_2=1.0$. Compared to the earlier case, one can see the enlargement of the radii of the annular structure. This is because of the fact that $J$ is negative here and when swarmalators are in nearby phases, their spatial attraction strength is reduced, which in turn increases the radii. Finally, when $K_2>K_2^*$, we see all the phases are synchronized and the annulus reaches an optimal state after which the structure remains invariant (see Fig.~\ref{annular-radii} (c)). We delineate the change of the radii as $K_2$ is varied in Fig.~\ref{annular-radii} (d). We find that both the outer ($R_{out}$) and inner ($R_{in}$) radii change as $K_2$ is increased until the critical point is reached at $K_2^*=1.5$. After this the radii do not depend on $K_2$. We can calculate this stationary value of the radii analytically which is present in Sec.~\ref{analytical}. We also illustrate the variation of $R$ in Fig.~\ref{annular-radii} (e).

\begin{figure}[t]
\includegraphics[width=1\columnwidth]{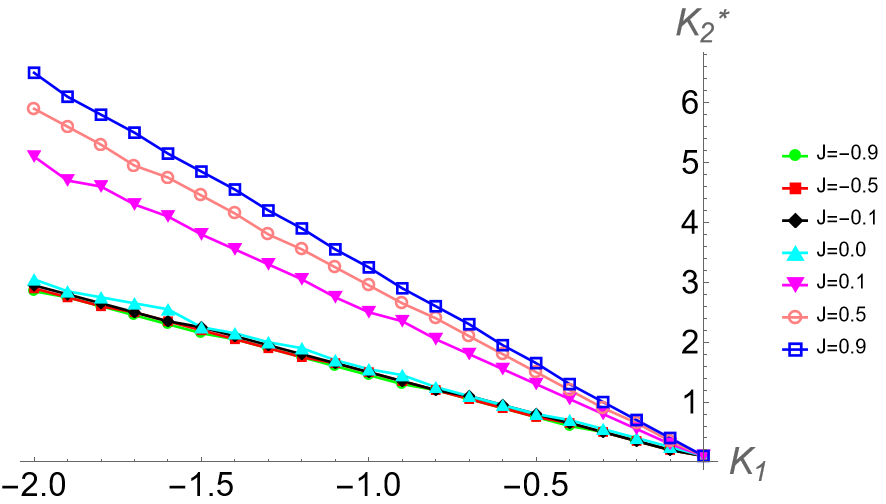}
\caption{The critical coupling strength $K_2^*$ as a function of $K_1$. A linear relation is observed. The investigation is performed with different $J$ values. For simulations we have used $N=500$ swarmalators. See main text for more details.}
\label{k2-critical}
\end{figure}

Moving on, we investigate the critical coupling strength $K_2^*$ beyond which synchrony is achieved and its relation with $K_1$. In Fig.~\ref{k2-critical}, we plot $K_2^*$ as we vary $K_1$ from $-2$ to $0$. We do this with different values of $J$. It is found that $K_2^*$ varies linearly with $K_1$. The indication is that the smaller the value of $K_1$ (larger is magnitude), the larger the value of $K_2$ should be to overcome the phase differences among swarmalators and achieve synchrony. We also see from Fig.~\ref{k2-critical} that for $J \le 0$ the critical $K_2$ values are almost same as the cyan ($J=0$), black ($J=-0.1$), red ($J=-0.5$), and green ($J=-0.9$) lines overlap. However, for positive $J$, a higher value of $K_2^*$ is needed for synchrony. Look at the magenta, pink, and blue lines for $J=0.1$, $J=0.5$, and $J=0.9$, respectively. To calculate $K_2^*$ numerically for a fixed $K_1$, we run simulations for $T=1000$ time units with step-size $dt=0.1$ with the RK4 method and check the values of $R$ at every $K_2$ value which is increased by $0.1$ at every step. The first $K_2$ value at which $R$ crosses 0.9 is then taken as $K_2^*$. We repeat this method for each value of $J$ mentioned in Fig.~\ref{k2-critical}.

\subsection{Effect of $J$}\label{effect-j}
We already know that $J$ controls the spatial attraction strength among the swarmalators. The term $1+J \cos(\theta_j-\theta_i)$ stands for the phase dependent spatial attraction strength among the swarmalators. For $J>0$ its maximum value is $1+J$ when swarmalators are in equal phase, and minimum value is $1-J$ when their phase difference is $\pi$. It means that when $J$ is positive, swarmalators in nearby phase will strongly attract each other spatially. On contrary, when $J<0$, the attraction is maximum at $\pi$ phase difference and minimum when they are in same phase. This means swarmalators in opposite phase will be strongly attracted to each other when $J$ is negative. 

Look at the top row of Fig.~\ref{states} where $J=-1.0$. When $K_1<0$ and $|K_2|<K_2^*$, swarmalators' phases are not fully synchronized. With negative $J$ swarmalators are less attracted towards each other that are in nearby phase and they tend to spread around the contrarian in an annular structure (see Fig.~\ref{states}(a)). When the phase couplings are absent, i.e., $K_1=K_2=0$, swarmalators stick to their initial phases that are uniformly distributed between $0$ to $2\pi$. The annular structure is deformed and the spatial structure resembles to the crescent moon (see Fig.~\ref{states}(b)). The annular restored with $K_1>0$ where sync is observed, as highlighted in Fig.~\ref{states}(c). For $J=0$, i.e., when the phases do not affect the spatial attraction, the annular structure is always maintained regardless of the value of $K_1$ and $K_2$. Look at the middle row of Fig.~\ref{states} where $J=0$. Finally, when $J>0$, nearby phase swarmalators attract one another and tend to stay nearby. As a result, the annular structure breaks down and crescent moon-like structure emerges (Fig.~\ref{states}(g) and (i)) except when $K_1=K_2=0$ where annular phase wave is observed (Fig.~\ref{states}(h)).

\subsection{Radii of the annular sync and async states}\label{analytical}
We study this in the continuum limit $N \rightarrow \infty$. Let $\rho(\mathbf{x},\theta,t)$ denote the probability density of a swarmalator at $\mathbf{x} \in \mathbb{R}^2$ having a phase $\theta$ at time $t$. The probability density follows the normalization condition, $\int \rho(\mathbf{x},\theta,t) d\mathbf{x} \; d \theta = 1$. It also follows the non-local integro-differential equation~\cite{chen2014minimal}
\begin{equation}
	\frac{\partial}{\partial t}\rho(\mathbf{x},\theta,t) + \nabla \cdot (\rho(\mathbf{x},\theta,t) v(\mathbf{x},\theta,t)) = 0, \label{continuity}
\end{equation}
where the velocity field $v(\mathbf{x},\theta,t) = (v_x,v_{\theta})$ is derived from Eqs.~\eqref{space}-\eqref{phase} as
\begin{align}
	v_x(\mathbf{x},\theta,t) = &\int \left(\frac{1}{|\mathbf{x}-\mathbf{x}'|^2}-(1+J \cos(\theta-\theta'))\right) \times\nonumber \\ &(\mathbf{x}-\mathbf{x}')\rho(\mathbf{x}',\theta',t)d\mathbf{x}' d\theta' - \mathbf{x} + \frac{\mathbf{x}}{|\mathbf{x}|^2},\\
	v_{\theta}(\mathbf{x},\theta,t) = &\int \frac{K_1 \sin(\theta'-\theta)}{|\mathbf{x}-\mathbf{x}'|}\rho(\mathbf{x}',\theta',t)d\mathbf{x}' d\theta' - K_2 \frac{\sin \theta}{\mathbf{x}}.
\end{align}
Eq.~\eqref{continuity} basically stands for the conservation of mass of the swarmalators as no swarmalator is created or destroyed at any point of time.
\begin{figure}[t]
\includegraphics[width=\columnwidth]{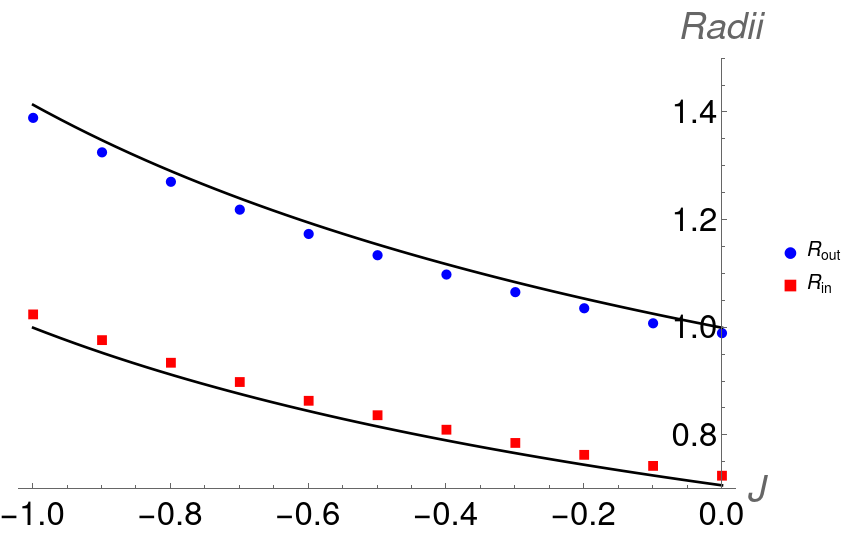}
\caption{Radii of the static sync state as functions of $J\le 0$. Black curves are analytical predictions given by Eqs.~\eqref{r-in}-\eqref{r-out}. Blue dots and red squares are the numerical values of $R_{out}$, and $R_{in}$, respectively. Here, we have taken $(T,dt,N)=(1000,0.1,500)$ for simulations. The other parameter values are $K_1=-K_2=1$.}
\label{sync-radii}
\end{figure}

\subsubsection{Radii of the static sync state for $J\le 0$}
The phases of the swarmalators are completely synchronized when $K_1>0$ or $|K_2|>K_2^*(K_1)$. In the sync state, since the phases are equal the effective attraction strength between the swarmalators become $1+J \cos(\theta_j-\theta_i) \rightarrow 1+J$. The static sync state is in annular shape when $J \le 0$. Suppose, $R_{out}$ and $R_{in}$ are the outer and inner radii of the annulus, respectively. The steady state density in the sync state is $\rho_0 = {1}/{\pi (R_{out}^2-R_{in}^2)}$ inside $\mathbb{A}$ and zero outside it, where $\mathbb{A}$ denotes the annulus. Using the methods described in Ref.~\cite{chen2014minimal}, we get
\begin{align}
	v(\mathbf{x}) &= \int_{\mathbb{A}} \left[ \frac{1}{|\mathbf{x}-\mathbf{x}'|^2}-(1+J)\right](\mathbf{x}-\mathbf{x}')\rho(\mathbf{x}',t) d\mathbf{x}' \nonumber \\ &- \mathbf{x} + \frac{\mathbf{x}}{|\mathbf{x}|^2} \nonumber\\
	&= \pi \rho_0 \mathbf{x}\left(1-\frac{R_{in}^2}{|\mathbf{x}|^2}\right)- (1+J)\pi \rho_0 \mathbf{x}(R_{out}^2-R_{in}^2) \nonumber \\ &- \mathbf{x} + \frac{\mathbf{x}}{|\mathbf{x}|^2},
	\label{velocity}
\end{align}
for $\mathbf{x} \in \mathbb{A}$. In the static sync state, $v(\mathbf{x})=0, \forall \mathbf{x} \in \mathbb{A}$. Now, if we choose $\mathbf{x}$ such that $|\mathbf{x}| = R_{in}$, then we get from Eq.~\eqref{velocity}
\begin{equation}
	R_{in} = \sqrt{\frac{1}{2+J}},
	\label{r-in}
\end{equation}
and by choosing $|\mathbf{x}| = R_{out}$, we find
\begin{equation}
	R_{out} = \sqrt{\frac{2}{2+J}}.
	\label{r-out}
\end{equation}
It is worth noting that $R_{out}$ and $R_{in}$ are independent of the values of $K_1$ and $K_2$ provided $K_1>0$ or $|K_2|>K_2^*(K_1)$ (when $K_1<0$). In Fig.~\ref{sync-radii}, we have shown that these analytical predictions match closely with our numerical results.

\begin{figure}[t]
\includegraphics[width=\columnwidth]{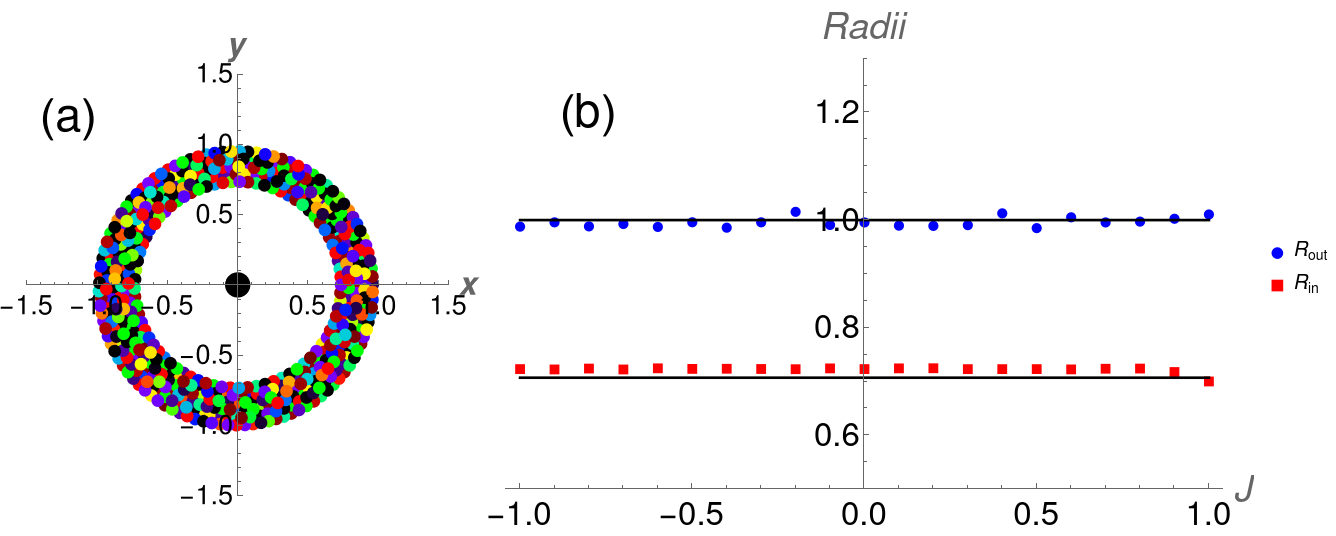}
\caption{Annular static async state. We have taken $K_1=-1,K_2=0$ here. Blue dots and red squares are the numerical values of $R_{out}$, and $R_{in}$, respectively. Black lines are analytical predictions given by Eq.~\eqref{async-rad}. We have used $(T,dt,N=1000,0.1,500)$ here.}
\label{async-radii}
\end{figure}

\begin{figure}[b]
\includegraphics[width=0.9\columnwidth]{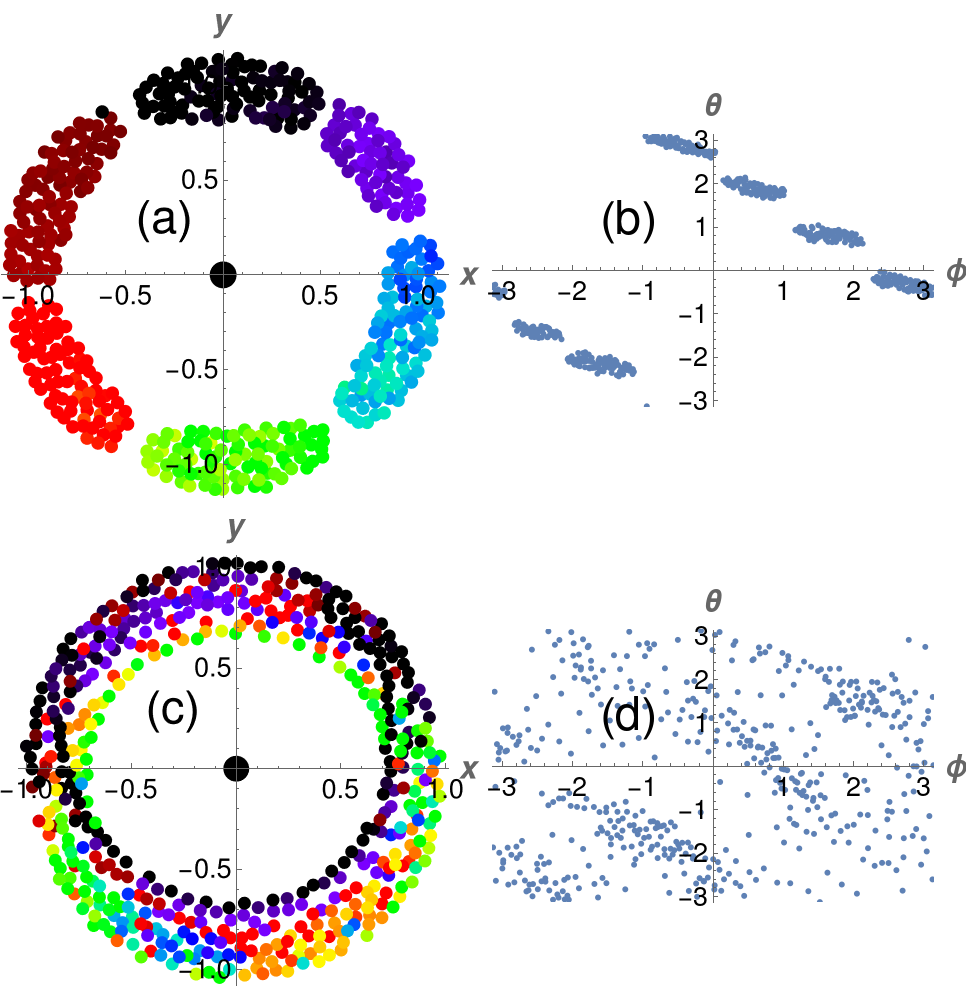}
\caption{Splintered phase wave and active phase wave states for $K_2=0.0$. In the top row, we show the splintered phase wave state with parameters $J=1.0$ and $K_1=-0.1$. (a) Swarmalators' positions are shown where they are colored according to their phases. (b) The relation between their spatial angles ($\phi$) and phases ($\theta$) are demonstrated. The bottom row conveys the same information for the active phase wave state with parameters $J=1.0$ and $K_1=-0.75$. We have used $(T, dt, N=1000, 0.1, 500)$ here.}
\label{active}
\end{figure}
\subsubsection{Radii of the static async state for $K_1<0$ and $K_2=0$}
In this case, the phases are desynchronized and as a result the cumulative spatial attraction force $1+J\cos(\theta_j-\theta_i)$ can be effectively taken 1. By using the same method as in the sync state, we find
\begin{equation}
	R_{in} = \sqrt{\frac{1}{2}}, \hspace{10pt} R_{out} = 1, \label{async-rad}
\end{equation}
in the async state. We have validated these expressions in Fig.~\ref{async-radii}. Note that, this only holds true when the phases are completely desynchronized and $R\approx 0$. If the value of $R$ lies between 0 and 1, then the radii of the annular state depends on the coupling strengths $K_1$ and $K_2$.

\begin{figure*}
    \includegraphics[width=1.7\columnwidth]{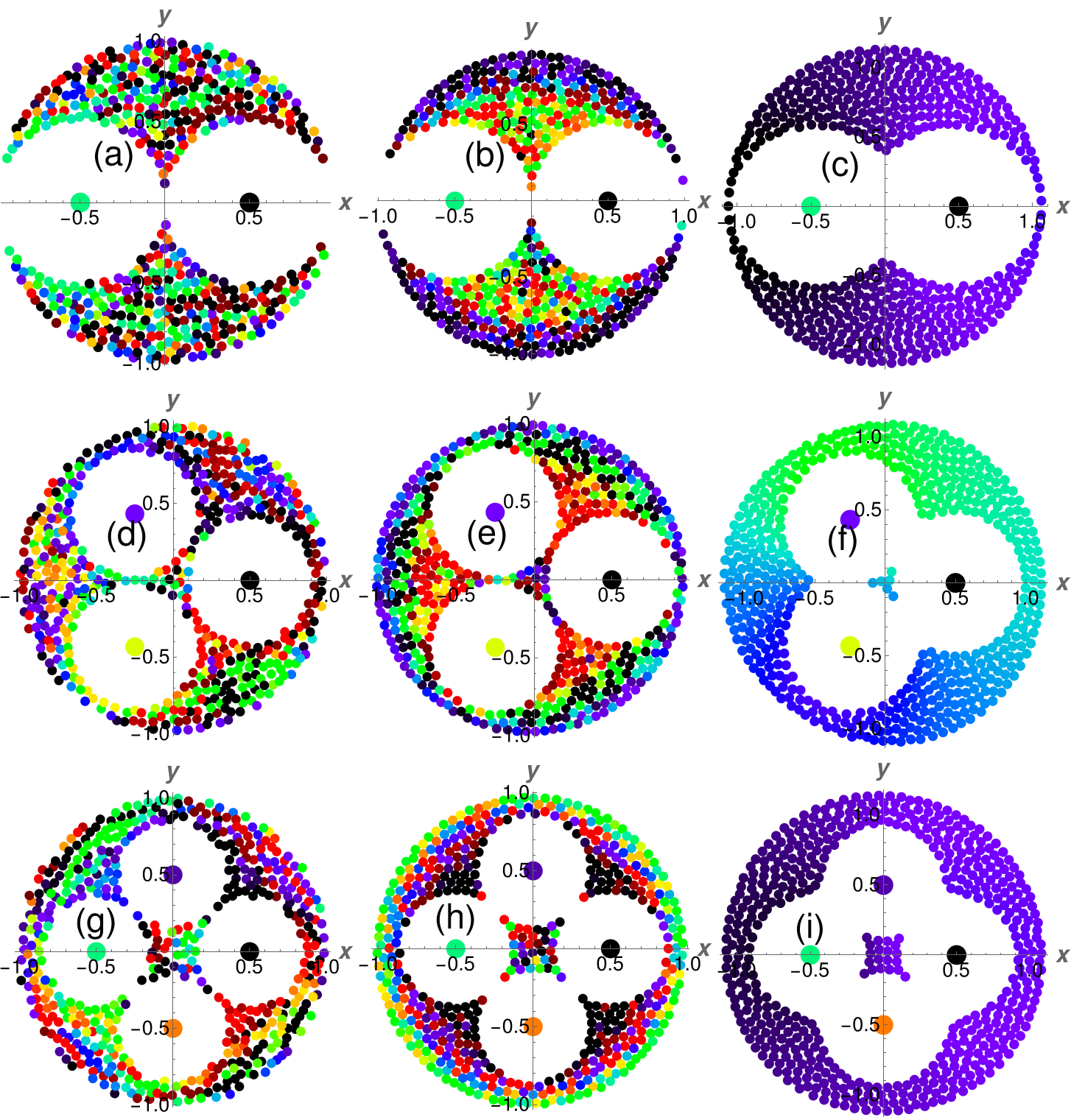}
    \caption{Emerging dynamics of the swarmalators of multiple contrarians. Swarmalators and contrarians are colored according to their phases. We run simulations with Eqs.~\eqref{eq3}-~\eqref{eq4} for $N=500$ swarmalators and $p=2,3,4$ contrarians where their positions and phases are determined from Eq.~\eqref{contrarian} with $r=0.5$. Top row: (a)-(c) stand for $p=2$, middle row: (d)-(f) show the results for $p=3$, and bottom row: (g)-(i) is for $p=4$. The coupling scheme is varied from one column to another where $K_1=-K_2=-0.5$, $K_1=-K_2=0$, and $K_1=-K_2=0.5$ in the first, second and third columns, respectively. We have fixed $J=-0.5$ for all the figures.}
    \label{multiple}
\end{figure*}

\subsection{Active states for $K_2=0$}
The simulation of the swarmalator model without the presence of contrarian in Ref.~\cite{o2017oscillators} revealed the existence of two \textit{active} states for negative phase coupling strength where the positions and phases of the swarmalators kept evolving over time. These states are called \textit{splintered phase wave} and \textit{active phase wave}. Here, we try to understand if these two active states exist in our model. For that, we consider that the contrarian's phase does not affect the swarmalators' phases, i.e., $K_2=0$. However, swarmalators' positions are influenced by the contrarian's position as before. By running simulations we have observed the emergence of the analogous versions of these two states that are depicted in Fig.~\ref{active}. In the splintered phase wave state, shown in Fig.~\ref{active}(a)-(b), swarmalators form disjoint clusters among them. They move inside those clusters and their phases also keep changing by small magnitude. In the active phase wave state (Fig.~\ref{active}(c)-(d)), swarmalators rotate inside the annulus. There exist two groups of swarmalators that exhibit counter-rotating motions on the annulus. Their phases also keep changing between $0$ to $2\pi$ in this state.

\section{Multiple contrarians, $p>1$}\label{mult}
So far, we have concentrated on the emerging dynamics of swarmalators in the presence of a single contrarian. Now, we want to explore the scenario when more than one contrarian are present in the system. The positions and phases of the contrarians are given by Eq.~\eqref{contrarian} where we choose $r_i=0.5$ for all $i \in \{1,2,\dots,p\}$. We run simulations for $p=2,3,4$ contrarians and delineate our findings in Fig.~\ref{multiple}. Here, we have fixed $J=-0.5$ that showcases significant dynamics. The emergent collective states are shown for two, three, and four contrarians in the top, middle, and bottom rows, respectively. The coupling strengths are chosen as $K_1=-K_2=-0.5$, $K_1=-K_2=0$, and $K_1=-K_2=0.5$ in the first, second, and third columns of Fig.~\ref{multiple}, respectively. We can see the emergence of new spatial patterns that depend on the number of contrarians and where they are positioned in the 2D plane.  The remarkable thing is the emergence of spatial structures with multiple vortices centered around the positions of the contrarians. These vortices all lie inside a disc. Looking deeply into the spatial structures presented in the top row of Fig.~\ref{multiple} for $p=2$, it is revealed that two vortices are formed surrounding the positions of the contrarians that are positioned at $(0.5,0)$, and $(-0.5, 0)$. Differently put, the contrarians maintain a nonzero distance from the swarmalators in each case. In Fig.~\ref{multiple}(a), $K_1=-K_2=-0.5$ and swarmalators' phases are desynchronized. In the absence of the coupling strengths ($K_1=-K_2=0$), swarmalators are rearranged depending on their phases, where the ones in nearby phases lie close to each other (Fig.~\ref{multiple}(b)). In (c), we have chosen $K_1=-K_2=0.5$, and the phases get synchronized as $K_1$ being positive. Similar scenarios are observed for $p=3$ in Figs.~\ref{multiple}(d)-(f), and $p=4$ in Figs.~\ref{multiple}(g)-(i).

\section{Discussions}\label{discussion}
We have proposed a swarmalator model under the influence of an external agent which we call a contrarian. We have found that the swarmalator dynamics is hugely impacted both by the contrarian's position and phase. Swarmalators feel spatial attraction and repulsion with the contrarian as they feel among themselves. This is the reason that the annular structure is prevalent in our study and the contrarian is positioned at the center of the annulus in all the cases. Similarly, swarmalators' phases are coupled to the contrarian's phase while they are also coupled among themselves. We find that swarmalators' phases get synchronized even with a negative coupling strength ($K_1<0$) when they are coupled to the contrarian's phase with a sufficiently large magnitude ($|K_2|>K_2^*$). Our model could be useful in studying the behavior of many natural and technological systems. We can describe this with an ecological example: consider a group of prey (for example, zebra) under the attack of a predator (for example, a lion). The natural tendency of the zebras is to avoid death by running away from the lion, and the lion chases them and tries to minimize its distance with the pack. These behaviors are modeled by the spatial attraction and repulsion functions in our system. Now consider a situation where the zebras act in unison (by synchronizing their phases) and surround the lion from all directions. The lion roars and tries to frighten the zebras by instilling fear among them (the phase interaction between the contrarian and the swarmalators). It then depends on how the zebras react to it and can eventually decide their fate. Other such examples can be found in predator attacks in fish schools, ant colonies, and so on. Other models can be tuned wisely to capture some of these phenomena.

While our research adds a new direction to swarmalators, it also puts forward a few questions that can be used as cues for future studies in this field. The common question to ask is what happens when one considers the time-varying nature of the contrarian in contrast to the static nature of it in our study.  We expect new spatial structures as the contrarians start to move in the 2D plane, as well as their phases also changing. We can also come across both static and active states: one where the contrarians keep moving and the one where they become stationary. One can also think about the spatial structure of the crescent moon like states where the annular structure disappears. Another interesting avenue would be to study the model for multiple contrarians, the initial results of which are delineated in Sec.~\ref{mult}. What are the radii of the vortices there and how are they positioned inside a disc? Some inspiration for solving this can be taken from Ref.~\cite{chen2013collective}. The contrarians can also be positioned at unequal distances from the origin by choosing different values for $r_i$ in Eqs.~\eqref{contrarian}. Our work can be extended by considering the positions of the swarmalators in three dimensions, or they can be taken in one dimension for solvability. It would be interesting to see how the dynamics unfolds in those cases. We think these are some very relevant questions to ask and finding answers to these requires further rigorous studies.

\bibliographystyle{apsrev}
%\bibliography{reference}

\end{document}